CONF-13-364-AD-APC-CD-TD# STATUS AND OPPORTUNITIES AT PROJECT X: A MULTI-MW FACILITY FOR INTENSITY FRONTIER RESEARCH*

S. D. Holmes, M. Kaducak, R. Kephart, I.Kourbanis, V. Lebedev, S. Mishra, S. Nagaitsev, N. Solyak, R. Tschirhart, Fermilab, Batavia, IL  60510, USA*Abstract*

Project X is a multi-megawatt proton facility being developed to support a world-leading program in Intensity Frontier physics at Fermilab. The facility will support programs in elementary particle and nuclear physics, with the potential for broader applications in materials and energy research. Project X is in the development stage with an R&D program focused on front end and superconducting rf acceleration technologies, and with design concepts for a staged implementation. This paper will review the status of the Project X conceptual development and the associated R&D programs.## PROJECT X MISSION AND GOALS

Project X is a high intensity proton facility that will support a world-leading U.S. program in Intensity Frontier physics over the next several decades. Project X is currently under development by Fermilab in collaboration with national and international partners. Project X will be unique in its ability to deliver, simultaneously, up to 6 MW of site-wide beam power to multiple experiments, at energies ranging from 235 MeV to 120 GeV, and with flexible and independently controlled beam time patterns. Project X will support a wide range of experiments utilizing neutrino, muon, kaon, nucleon, and atomic probes [1,2]. In addition, Project X will lay the foundation for the long-term development of a Neutrino Factory and/or Muon Collider.

## PROJECT X REFERENCE DESIGN

A complete design concept, designated the Project X Reference Design [3], has been established supporting the above-described mission in an innovative and flexible manner. The Reference Design is based on a 3-GeV superconducting (SC) continuous wave (CW) linac, followed by an 8-GeV superconducting pulsed linac, and improvements to the Main Injector complex at Fermilab. The primary elements of the Reference Design are:

- An $H^-$ front end based on a 162.5 MHz RFQ delivering 5 mA of beam current at 2.1 MeV into a medium energy beam transport (MEBT) line containing a wideband, bunch-by-bunch, chopper capable of accepting or rejecting bunches in arbitrary patterns and leaving 1 mA average (averaged over ~1 μsec) current for further acceleration;
- A 3-GeV SC linac operating in CW mode and capable of accelerating an average beam current of 1 mA;
- A 3 to 8-GeV pulsed SC linac capable of accelerating a peak current of 1 mA with a duty factor of 4.3%;
- Rf splitters that can deliver beams at 1 and 3 GeV to at least three experimental areas;
- Experimental facilities to support experiments at 1 and 3 GeV;
- Modifications to the Main Injector (MI) and Recycler complex required to support 2-MW operations in the energy range of 60-to-120 GeV;
- All interconnecting beam lines.

## STAGING STRATEGY

Fiscal considerations have motivated development of a staging plan for Project X, consistent with the following principles:

- The plan should provide compelling physics opportunities at each stage;
- Each stage should cost substantially less than $1B;
- Each stage should utilize existing infrastructure to the extent possible;
- The construction of each stage should minimize interruptions to the on-going research program;
- The full Reference Design capabilities should be achieved at end of the final stage.

The Reference Design has been developed consistent with these principles – the associated layout on the Fermilab site is shown in Figure 1.

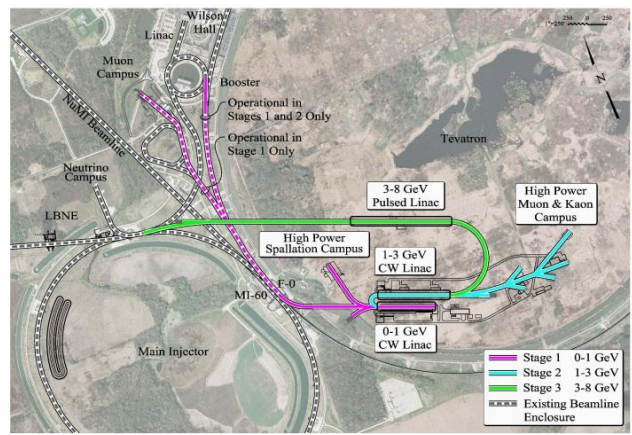

Figure 1: Project X Reference Design Stages

Stage 1 is based on the 1-GeV CW linac. The 1-mA beam can be directed either to the existing 8-GeV Booster, the Muon Campus (currently under construction), or to a newly constructed High Power Spallation Campus. Stage 1 retires the existing 400-MeV linac and raises the Booster injection energy to 1 GeV,

_________________________________________
* Work supported by the Fermi Research Alliance under U.S. Department of Energy contract number DE-AC02-07CH11359

allowing a 50% increase in beam intensity. Stage 1 can simultaneously provide 1.2 MW to the long baseline neutrino target at 120 GeV, 80 kW to the Mu2e experiment [4], and 900 kW of beam power to the new Spallation Campus.

In Stage 2 the CW linac is extended to 3 GeV. The average beam current in the 1-GeV linac is increased to 2 mA with 1 mA available for the Booster and the Spallation Campus, and 1 mA passed into the 3-GeV linac for delivery to a newly constructed High Power Muon and Kaon Campus.

Stage 3 completes the Reference Design with the 8-GeV pulsed linac replacing the Booster as the injector into the Main Injector complex. The pulsed linac delivers a 1-mA peak beam current with a 4.3% duty factor (4.3 ms × 10 Hz). At Stage 3 the Main Injector can deliver 2.3 MW of beam power at any energy between 60-120 GeV. In addition, beam power is available for an 8 GeV program ranging from 0-170 kW, depending on the operating energy of the Main Injector.

The capabilities of Project X at each stage are given in Table 1.

Table 1: Project X Performance by Stage

|  | Stage 1 | Stage 2 | Stage 3 |
| --- | --- | --- | --- |
| 1 GeV Beam Power | 0.9 MW | 1 MW | 1 MW |
| 3 GeV Beam Power | --- | 3 MW | 3 MW |
| 8 GeV Beam Power | 110 kW | 110 kW | 350 kW |
| 120 GeV Beam Power | 1.2 MW | 1.2 MW | 2.3 MW |

## R&D PROGRAM

The Reference Design provides the context for the Project X R&D program. The purpose of the R&D program is to mitigate technical and cost risks. The primary risks associated with Project X are identified as:

- Front End: The front end must provide an average beam current of 1-2 mA, with a peak current of 5-10 mA, with the beam assigned in a programmable manner into the linac rf buckets.
- H$^-$ injection: All stages of Project X require injection via stripping over hundreds of turns. Space-charge mitigation requires foil painting. Other considerations include foil heating, and losses generated by partial stripping and/or or multiple scattering.
- High Intensity Main Injector and Recycler operations: Operations at intensities a factor of three beyond current experience require upgrades to the rf systems and control of a variety of beam instabilities.
- High Power Target Facilities: Target facilities for producing both secondary particle beams and spallation neutrons must accommodate >1 MW of incident beam power.
- Superconducting rf: A variety of cavities operating at different frequencies and with different geometric betas are required.

The R&D program as structured could support a Project X construction start in 2018.

### Project X Injector Experiment (PXIE)

PXIE is an integrated front-end systems test, and is the centerpiece of the Project X R&D program [5]. PXIE will validate the concept for the Project X front end, thereby minimizing the primary technical risk element within the Reference Design. The PXIE layout is shown in Figure 2.

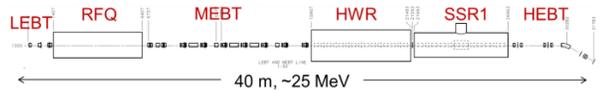

Figure 2: PXIE Layout

PXIE will operate with the full Project X design parameters. These include delivering 5 mA average current from the RFQ, followed by 80% chopping within the MEBT for an average current delivered to the SC accelerating section of 1 mA. Two SC acceleration modules (HWR and SSR1) are expected to provide efficient acceleration with minimal emittance dilution to ~25 MeV. PXIE will utilize components constructed to Project X specifications wherever possible. This will provide an opportunity to re-utilize selected PXIE components within Project X.

PXIE is being undertaken by a collaboration of Fermilab with ANL, LBNL, ORNL/SNS, and four Indian institutions (BARC, IUAC, RRCAT, VECC).

### Superconducting RF

The Project X linacs utilize modern SC radiofrequency (srf) technology as it has been developed over the last few decades. This technology is considered mature, and does not represent a significant technical risk within Project X; it is, however, a primary cost driver.

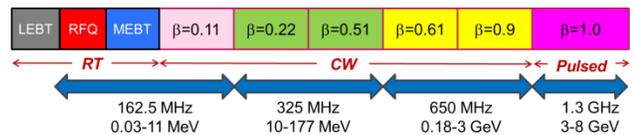

Figure 3: Project X Linac Technology Map

Figure 3 displays the technology layout of the Project X linacs. A room temperature (RT) front end is followed immediately by superconducting accelerating modules extending from 2.1 MeV to 8 GeV. The linac operates in CW mode up to 3 GeV and in pulsed mode thereafter. Operation of a CW linac places a strong premium on achieving high $Q_0$ in order to reduce both operating costs and the capital investment in the cryogenic support systems [6]. A robust srf R&D program has been underway for several years at Fermilab in collaboration with national and international partners. This program originated as an initiative to develop 1300 MHz accelerating structures for the International Linear Collider [7] in the mid-2000s. The significant investment in ILC technologies made it natural to base the Project X design on sub-harmonics of 1300 MHz.

The status of srf development for Project X is summarized in Table 2. HWR and SSR signify half-wave and single-spoke resonators; LE650 and HE650 are both 5-cell elliptical cavities; while ILC is a 9-cell elliptical cavity (operating at lower gradient, but higher $Q_0$ than ILC) Substantial progress has been made on all cavity types. As an example Figure 3 shows test results from six SSR1 cavities fabricated in industry, all of which meet the Project X performance specification and are planned to be used in PXIE and later in the Project X front end.

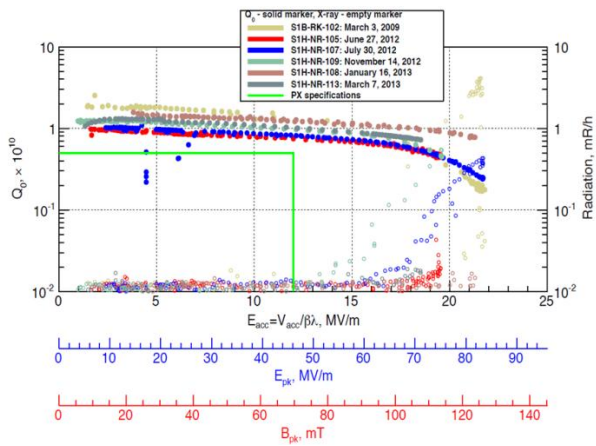

Figure 4: Dependence of $Q_0$ on accelerating gradient for SSR1 bare cavities at 2K. The green lines indicate the Project X specification.

## SUMMARY

Project X is central to the strategy for future development of the Fermilab accelerator complex. A Reference Design has been developed that will support a world-leading program in neutrinos and other rare processes over the coming decades, and will provide a platform for future muon-based facilities. A staging strategy has been developed that provides compelling physics opportunities at each stage. An R&D program is underway with particular emphasis on front end and srf development. Project X could be ready to initiate construction in the second half of the current decade.

Table 2: Project X SRF Accelerating Cavity Requirements and Status.

| | Frequency (MHz) | Number Required | Gradient (MV/m) | $Q_0$@2K ($10^{10}$) | Collaborator | Status |
|---|---|---|---|---|---|---|
| HWR ($\beta_{opt}$=0.11) | 162.5 | 8 | 8.2 | 0.5 | ANL | Cavity and cryomodule design complete; prototype cavity on order |
| SSR1 ($\beta_{opt}$=0.22) | 325 | 16 | 10 | 0.5 | India | Ten cavities fabricated; 5/6 tested and meet spec; two cavities in fabrication at IUAC |
| SSR2 ($\beta_{opt}$=0.51) | 325 | 35 | 11.2 | 1.2 | India | Cavity electromagnetic and mechanical designs complete |
| LE650 ($\beta_G$=0.61) | 650 | 30 | 16.5 | 1.5 | India JLab | Cavity electromagnetic and mechanical designs complete |
| HE650 ($\beta_G$=0.9) | 650 | 162 | 17 | 2.0 | India | Cavity electromagnetic and mechanical designs complete; single cells meet spec; 5-cell protoypes received |
| ILC ($\beta_G$=1.0) | 1300 | 224 | 17 | 2.0 | DESY KEK | >40 cavities tested; two prototype cryomodules complete and tested to PX spec. (ILC program). |